\documentclass[twocolumn,letterpaper,byrevtex,showkeys]{revtex4}
\usepackage{graphicx}
\usepackage{amsmath}
\usepackage{amsfonts}
\usepackage{amssymb}

\begin{document}

\begin{abstract}
The results of an extensive operation of a Vacuum Spark plasma
using Titanium electrodes in a 120 ns 150 kA discharge are
presented.  The hot spots are found to form with a regular spacing
in a zippering Z-pinch plasma, which forms close to the cathode
and extends to approximately two thirds of the anode separation
over a period of a few ns.  The axis of the discharge is well
defined by an initial plasma from a Nd:YAG laser focussed onto the
cathode electrode surface.  The statistics of the formation of the
hot spots are given for the life of one anode electrode.  Between
one and three hotspots form and the favored positions are at 1.5
and 3.0 mm from the cathode and the strongest emission, as
observed in a filtered X-ray pinhole camera, comes from the hot
spot closest to the cathode.  The emission spectra resolved
between 50 and 350 \AA\, shows a wide range of Ti ionization which
allows the temperatures of the anode blow off plasma, the Z-pinch
and the hot spot plasma to be distinguished.  These results are
compared with filtered PIN diode signals and filtered pinhole
images.

\end{abstract}

\title{Reproducibility of a Titanium Plasma Vacuum Spark Discharge}

\author{Edmund S. Wyndham}
\email[E-mail:\hspace{1mm}]{ewyndham@fis.puc.cl}

\author{Mario Favre}
\author{Hern\'an Chuaqui}
\author{Ana M. Le\~nero}
\author{Jorge S. D\'\i az}
\affiliation{Pontificia Universidad Cat\'olica de Chile,
Departamento de F\'isica, Casilla 306, Santiago, Chile.}

\author{Peter Choi}
\affiliation{EPPRA sas, 16 Avenue du Qu\'ebec, 706 SILIC,
Courtaboeuf 91961, France}

\date{December 31, 2004}
\maketitle

\section{INTRODUCTION}

The Vacuum Spark \cite{01_L.Cohen} has long attracted interest as
a rich source of plasma physics phenomena which have been
thoroughly reviewed \cite{02_E.D.Korop,03_K.N. Koshelev}.  The
most researched are the hot spots, often at extreme conditions of
density and temperature.  While the Vacuum Spark has provided much
spectral information of highly charged metallic ions, its
development as a reliable source for advanced technological
applications has been less successful. Vacuum Spark discharges
have been realized with various electrode shapes as well as with
slow low voltage capacitor banks and medium power pulsed power
coaxial line generators.  Various trigger schemes have been tried
to obtain better and more consistent operation of the discharge. A
trigger spark behind the cathode has been generated electrically
or by a focussed laser, a laser has been focussed onto the front
surface of the anode and cathode. The hot spots have been observed
in the anode blow-off plasma and as necking instabilities in the
Z-pinch plasma which forms between the cathode and anode, beyond
the blow-off plasma.  Particular interest in the hotspots arises
from the fact that, due to the high Z of the ions charge state,
they exhibit the typical characteristics of radiative collapse a
the Z-pinch plasma.

The present work form part of a series of results
\cite{04_E.Wyndham} in which a Vacuum Spark configuration using a
small pulsed power generator firsts presented by Zakharov et al.
\cite{05_Zakharov} has been developed.  The salient features of
this configuration are that a relatively low energy laser pulse is
focussed onto the front cathode surface though a perforated
conical anode, which has a flat surface, rather than the rounded
or pointed extreme commonly used.  The electrical driver is a
pulsed power 2 Ohm coaxial line.  However, an important difference
from previous work is that the hybrid mode \cite{04_E.Wyndham} of
operation has been found to be preferred to the normal switched
line operation. In the hybrid mode the line gap is shorted so that
the line charging ramp voltage is also applied to the Vacuum
Spark.  The Vacuum Spark itself switches the line. This results in
a slightly longer applied current pulse with a lower maximum rate
of current rise of $1.5 \cdot 10^{12}$ A/s. Close to maximum
current a Z-pinch is found to zipper upwards from cathode towards
the anode, giving rise to the formation of a string from one to
three hot-spots. This work presents some statistical observations
of the frequency and the position the hotspots formed during the
useful life of one anode cone electrode.  The spectrum of a source
for lithographic applications is of essential importance and we
present a soft X-ray spectrum obtained with Ti electrodes.  The
relative timing between the X-Ray emission from the anode plasma
and the Z-pinch column is also presented.

\section{EXPERIMENTAL CONSIDERATIONS}

The principal details of the operation of the Vacuum Spark have
been described elsewhere \cite{04_E.Wyndham}. The observations of
the hot spot positions in the Z-pinch were taken with an electrode
separation of 7.75 mm.  The generator was operated in the hybrid
mode.  The laser pre-ionizing pulse was focussed at 0.3 J onto the
cathode surface approximately 700 ns before the application of the
generator.  The hot spots were observed in a time integrated
quadruple pinhole camera using four filters \cite{06_Chuaqui}. The
images were recorded on Kodak DEF film and the spatial resolution
was 200 $\mu$m for the pair of pinholes with softer filters, and
400 $\mu$m for the pair with harder filtering.  The softest of the
four filters was 1.5 $\mu$m Al + 0.4 $\mu$m Zn and the hardest was
50 $\mu$m Be, whose transmission window may be taken as cutting
off ($<$1\% transmission) in this experiment at 13 \AA.  The Zn
was added to the Al in order to eliminate the long wavelength
window of Al from about 170 to 350 \AA, while still preserving a
short wavelength pass band to 25 \AA.  The soft X-ray spectra were
taken with a compact grazing incidence Rowland circle spectrometer
with a 600 lpi grating and a useful spectral window of from 50 to
350 \AA. The image was recorded on film using a multi-channel
plate intensifier, with a 3 ns gating time.

\begin{figure}[ht]
\includegraphics[width=0.4\textwidth]{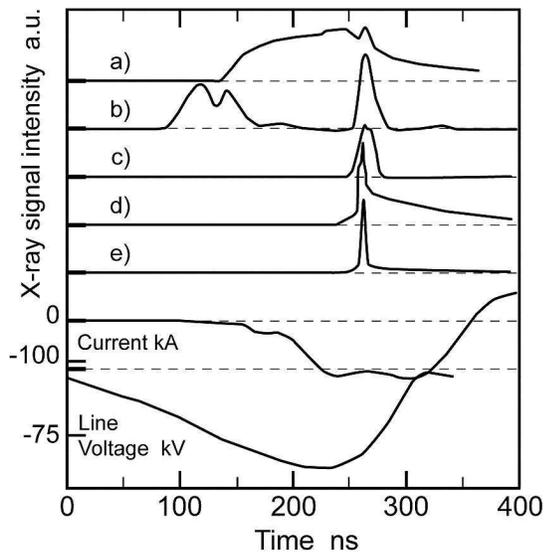}
\caption{X-ray emission from the whole discharge operated in the
hybrid mode, a) and b), and from half of the discharge closest to
the cathode, c) to e). The current and voltage on the line
transfer section are the two lower traces. Filters are: a) Al 3
$\mu$m, b) Ag 3 $\mu$m, c) Mylar 5 $\mu$m + Al 0,28 $\mu$m, d) Ti
1.2 $\mu$m and e) Be 34 $\mu$m.} \label{figure1}
\end{figure}

In figure 1 the time resolved X-ray emission is shown from the
whole volume of the discharge and from the 5 mm closest to the
cathode, where the Z-pinch column is observed to form.  The
discharge holds the full line charging voltage for approximately
150 ns before the current exceeds a few kA.  The 3 $\mu$m Ag
filter shows the early beam target emission from the anode, as the
current builds rather slowly to a few kA.  Slightly later the 3
$\mu$m Al filter registers emission from the plasma forming from
the ablated material of the anode.   The 3 $\mu$m Ag filter also
records the hotter plasma from appreciable beam target emission
from the anode and the anode plasma as well as from the hotspots
of the Z-pinch.  In a very crude approximation it may be said that
in this quite hard filter the emitted energies are approximately
equal.  The plasma emission from the hot spots occurs at an
inflection of the peak current of approximately 150 kA. The
inflection is an artefact of the generator.  The softest of the
three filter signals that observe only the Z-pinch, 1.2 $\mu$m Ti,
shows the 15 ns period in which the Z-pinch channel is formed. The
hot spot signal occurs in the halfway through this filter signal,
as is seen from the other two filters.  The hardest filter, 34  m
Be, has the shortest signal and records principally Ti XIX and
higher ionization stages associated with the hotspot formation. It
is worth stressing that a multi keV component of emission has
never been observed in this configuration of vacuum spark if hot
spots are formed with the characteristic energy signature of this
figure.

\begin{figure}[ht]
\includegraphics[width=0.45\textwidth]{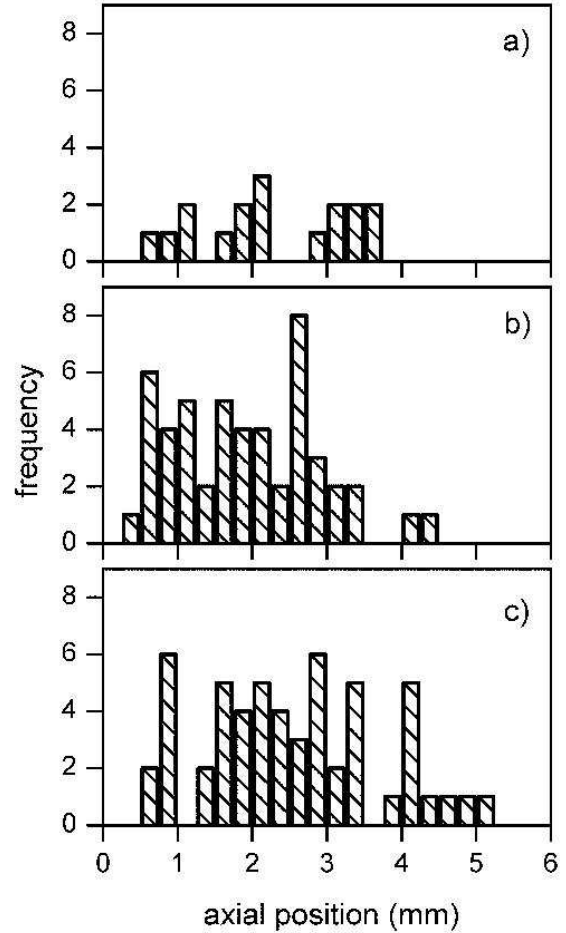}
\caption{Histograms of the axial position measured from the
cathode of the hot spots of 59 shots, when one hot spot is
observed , upper graph, when two hot spots are present, middle
graph and for three hot spots, lower graph.} \label{figure2}
\end{figure}

\begin{figure}[ht]
\includegraphics[width=0.45\textwidth]{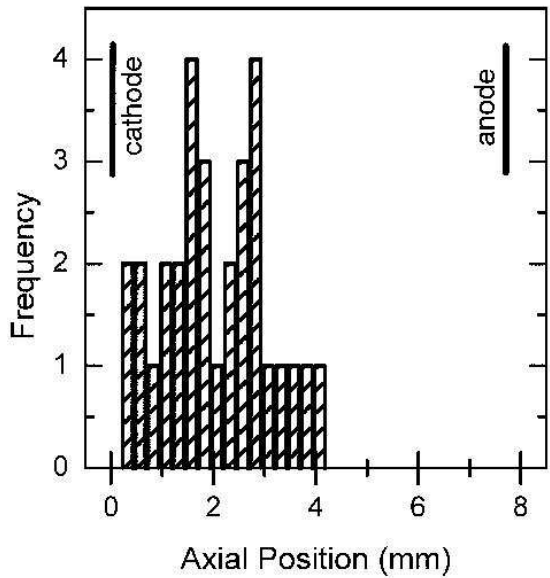}
\caption{Histogram showing the axial position of the brightest hot
spot for the same series of discharges as figure 2.}
\label{figure3}
\end{figure}

The statistical properties of the hot spots are recorded for a
sample of 59 shots taken on one pair of electrodes.  The useful
life of the electrodes is approximately 150 shots.  Of these 59
shots, 18 had one hotspot, 26 had two and 15 had three.  It should
be noted that all shots produce hotspots.  The only requirement
for consistent operation over a substantial number of shots is
that the laser focus must not be too tight in order to avoid
pitting in the cathode.  By defocussing the laser to a focal spot
of 300 $\mu$m diameter in excess of 100 shots can be obtained
before realignment is required.  The distribution of the axial
position measured from the cathode for these hot spots is shown in
a series of three histograms in figures 2a, 2b and 2c, for one two
and three hotspots respectively.  The spacing of the electrodes
was 7.7 mm, which was found to be the optimum spacing for the
production of hot spots. The most favored position for a hot spot
is between 2.6 and 3.2 mm with a total of 29 hot spots.  The next
most favored position is between 0.6 and 1.2 mm from the cathode,
with a total of 27 hot spots.  The frequency of hot spots per unit
axial length rapidly diminishes from 4 mm; as between 4 to 4.6 mm
only 9 are observed.  The position of the hottest and also the
most intense spot is plotted in figure 3.  The hotspot is nearly
always the only one visible in the 50 $\mu$m Be filtered pinhole
image.  This histogram corresponds to the 31 shots where two or
more hot spots were observed.  On comparing figures 3 and 4, it
may be easily seen that there is a correlation between the
position of greatest frequency of occurrence and the most intense
hotspot.

A spectrum from the whole plasma taken with the grazing incidence
spectrometer is shown in figure 4.  Within the time resolution of
the micro-channel gating, the recorded spectrum corresponds to the
time of hot spot formation.  As the spectrum covers a wide range,
it is shown in two halves to appreciate the smaller scale details.
A wide range of ionization stages is found indicating a plasma
with regions of widely differing temperature.  The ionization
stages cover Ti V to Ti XX, in three or four groups which may be
associated with different plasmas as will be discussed as follows.
The second order transitions are indicated with a down-pointing
arrow.  The four Ti XX lines are second order transitions, the
corresponding first order lines are seen, but are close to the
lower useful limit of the grating and are not indicated in the
figure.  The presence of this ionization stage is to be expected
from the presence of an image on the 50 $\mu$m Be filter and
corresponds to the hotspot plasma previously estimated  from
filtered PIN diode and filtered pinhole images at about 700 eV
\cite{06_Chuaqui}.  The following group of ionization stages that
may be identified is from Ti XVI to Ti XVIII.  For this group
there are a large number of lines that may be identified.  Their
intensities are in accordance with available transition
probabilities in the NIST database \cite{07_NIST}.  The lines
indicated are those whose transition probability value $A_{ki}$ is
greater than approximately $4\cdot 10^9$ s$^1$, when the value is
available, or the brightest observed lines in the notation of
Kelly database \cite{08_Kelly}. Ti XV is found to be all but
absent, the only unambiguous observation is at 147.4 \AA, while
the equally probable transition at 115.0 \AA\, is not
unambiguously identifiable. On the other hand Ti XIII and Ti XIV
are well represented.  Ti XII is most strongly represented by two
of the stage's transitions at 116.5 and 140.3 \AA, while others
with a high transition probability at 90.5, 108.1 and 109.1 \AA\,
are weak. In comparison it would appear that Ti XIII and Ti XIV
are more abundant.  Evidence of lower temperature plasma is found
from the ionization stages from the representative lines
identified from Ti VI to Ti XI.

\begin{figure}[ht]
\includegraphics[width=0.5\textwidth]{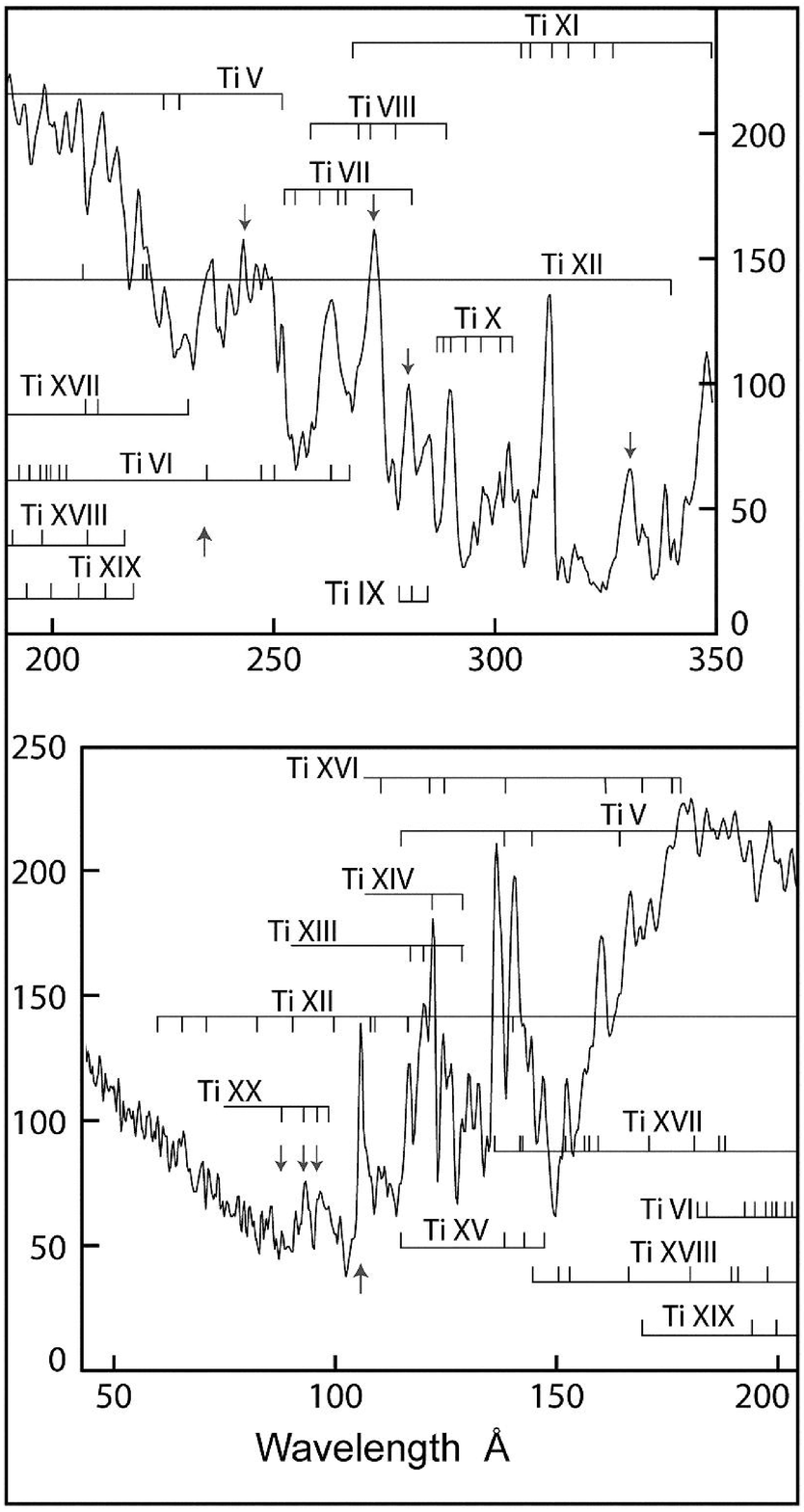}
\caption{Soft X-ray spectrum for one representative shot from 50
to 350 \AA. The single spectrum is divided in two sections for
clarity. The most probable transitions for the Ti ionization
stages that coincide with observed lines are shown. The downwards
directed arrows denote second order lines and the upwards directed
lines are anomalies in the Ti spectra. The vertical intensity
scale is in arbitrary units and is not linear.} \label{figure4}
\end{figure}

An anomaly is presented in the form of two lines, one of them very
prominent at 106.5 \AA, indicated by an upwards pointing arrow.
While there are Ti XV, XVI and XVII lines tabulated within the
resolution of the spectrograph at 106.5 \AA, their transition
probabilities are an order of magnitude below the criteria of the
other lines identified of these ionization stages. Lines from
possible contaminants, such as Cu, Zn, C, Si and F may all be
discounted. There are many extremely fast dynamic processes
associated with the formation and dissipation of a hot spot,
including plasma jets and charged particle beams, which might
affect the transition probabilities giving rise to energy level
shifts or polarization enhancements \cite{09_HR Griem}.  A more
detailed analysis of this unidentified emission lines is beyond
the present investigation.

\section{DISCUSSION}

In common with other fast vacuum discharge work referred to above,
the hot spots are seen to form a Z-pinch which "zippers" up within
a few ns starting from the position defined by the incident laser
on the cathode.  This formation was seen first in interferograms
\cite{04_E.Wyndham}, then in visible streak photography
\cite{06_Chuaqui} and later in soft X-ray frames
\cite{10_Chuaqui}.  In soft filtered pinhole images of the Z-pinch
column \cite{06_Chuaqui}, a plasma volume which extends two thirds
of the distance to the anode is seen to be nearly continuous, with
slight bulges where the hot spots are observed. In the hardest
filter, which is transparent to emission from Li-like and higher
ionization stages, a much more compact and irregular axially
elongated plasma volume is seen.  In the optical interferometric
images the anode plasma is seen to ``boil'' off the anode with too
many shift fringes or even optically dense for any estimate of the
electron density.  However soft X-ray emission from the plasma is
barely seen in micro-channel plate frames even in 3 $\mu$m
filtered images.  The anode itself is usually quite a bright
emitter, although an order of magnitude less than the Z-pinch,
except in the case where the plasma density is too low in the
pinch region and very intense e-beams arise.  On relating these
observations to the spectrum presented, it is reasonable to infer
that the ionization stages from Ti V to XII are associated with
the plasma boiling off from the anode.  These ionization stages do
not have any lines that would appear in images filtered with 1.5
or 3.0 $\mu$m of Al.  The plasma temperature may be obtained from
the collisional radiation code, FLY \cite{11_Lee}. The stages from
Ti XIX to Ti XX are associated with the hottest plasmas of the
hotspots. Li-like Ti has many lines in the pass-band of the Be
filter. Some of the lower energy transitions are seen in the
spectrum.  The most intense lines of the wavelength range of
spectrum are those corresponding to the Ti XIII to Ti XVIII
species.  Such a wide range of stages is not expected from a
single temperature plasma but rather the species are consistent
with a range of temperatures from 200 to 500 eV.  All of these
stages have lines that may be seen in the softer filters used in
the pinhole image camera. For these pinholes and especially the
case of the 2.0 $\mu$m Ti filter, with its L-shell pass band
between 28 and 35 \AA, the 1 mm diameter Z-pinch column is the
principal feature.  Hence it may be inferred that this plasma
occupies this range of ionization stages.  A wide variety of
ionization stages has been found in earlier work
\cite{02_E.D.Korop}, where Fe XVIII to XXVI were recorded in the
same discharge.

At present most effort in intense X-ray sources for lithography is
centered on the wavelengths where multilayer mirrors are
available, that is between 120 and 150 \AA.  In this range we
observe a number of comparatively intense lines of Ti XIV and
XVII.  No absolute X-ray measurements of the soft X-ray emission
is available in this work.  Absolute measurements have been
obtained using rather low efficiency multilayer mirrors at much
shorter wavelengths, centered on 8 and 24 \AA\,
\cite{12_M.Hebach}.

The reproducibility of hotspot formation has been studied in fast
capacitor bank driven vacuum sparks.  In an early work
\cite{13_Wong} a laser beam was focussed onto the anode electrode
and the probability of observing the hotspot in the plasma formed
from electrode ablation under different conditions of electrode
separation was observed.  The position of the hotspot was well
defined but the probability of occurrence was never greater than
80\%.  This work is the first to present a statistical analysis
over a large number of shots.  More recent laser triggered work in
the vacuum spark \cite{14_Ohzu} has concentrated on hard X-ray
emission. Other recent work \cite{15_N.Georgescu} has studied
various methods of electrical triggering of fast capacitor bank
vacuum sparks, with characteristic X-ray energies of between 3 and
40 keV.  Other work on reliable vacuum spark operation includes
the generation of a sliding spark on a PTFE sleeve covering part
of the cathode \cite{16_Rout}. The system is called passive, as
part of the applied generator voltage is resistively fed to a ring
electrode on the PTFE sleeve to initiate the sliding spark.

In this context, the present work may be the first to present
statistical results of a fast pulsed power driven vacuum spark.
Whereas there are a number of favoured axial positions, where, if
for example a 0.5 mm axial bin length is considered, there is a
100\% probability of finding a hotspot.  This hot spot is,
however, only an approximately 25\% probability of finding the
hottest hotspot even at the most favoured positions of 2 and 3 mm
from the cathode.  However modification of the experimental
assembly is possible so that the X-ray energy is extracted through
the hollow anode, in which case the entire integrated axial X-ray
emission would be available.  In this case the laser triggering
onto the cathode surface would come in obliquely, at approximately
$30^\circ$ from the normal.  As the Z-pinch column is well
stabilized on its axis by the laser pre-ionizing spark and is
observed to be well formed, with only $m=0$ instabilities, end-on
extraction of X-ray energy should present an order of magnitude
improvement in brightness over any radial energy extraction.

Observations of the Si pin diode signals over many shots indicate
a less than 35\% shot to shot variation for the softest filter
used, sensitive out to approximately 25\AA, whereas the variation
of harder filters sensitive to Li-like and He-like transitions
shows a 60\% shot to shot variation.  The temporal behaviour of
the X-ray pulse and the voltage and current waveforms are very
reproducible.

\section{CONCLUSIONS}

The vacuum spark driven by a small pulsed power generator and
triggered by a laser spark on the cathode compares very favourably
with lower voltage fast capacitor bank drivers.  In particular the
hotspots are formed within a rather cooler and well formed Z
pinch.  The characteristic line emission from this pinch is from
Ti XIII to XVIII ionization stages.  A statistical analysis of the
$m=0$ hotspot special occurrence indicates favoured positions
which also coincide with the brightest hotspot, which has very
significant Li-like emission.  The anode plasma is very dense and
considerably cooler.  No hotspots are ever formed in the ablated
anode plasma.  Whereas extraction of the soft X-ray energy in a
radial direction is possible, considering a 0.5 mm axial segment,
the experimental scheme is susceptible to important improvement
using axial energy extraction.

\section{ACKNOWLEDGEMENTS}

The authors wish to express their gratefulness to the Chilean
government research fund FONDECYT 1030968 and important capital
equipment investment from the Andes Foundation, grant C-13768.

\end{document}